\documentclass[a4paper,fleqn,twocolumn]{aastex63}

\usepackage{amsmath}	
\usepackage{amssymb}	
\usepackage{array}	
\usepackage{hyperref}

\newcommand{\Soltan}{So\l tan's~}

\defcitealias{nu2gc}{M16}
\defcitealias{YT02}{YT02}
\graphicspath{{figures/}}

\received{June 1, 2019}
\revised{January 10, 2019}
\accepted{\today}
\submitjournal{AJ}

\shorttitle{Super-Eddington growth of SMBHs}
\shortauthors{Shirakata et al.}

\begin{document}

\title{Revisiting \Soltan argument based on a semi-analytical model for galaxy and black hole evolutiton}

\author{Hikari Shirakata}
\affiliation{Department of Cosmosciences, Graduate School of Science, Hokkaido University, N10 W8, Kitaku, Sapporo, 060-0810, Japan}
\affiliation{The Technical Research Center, Tadano Ltd., 2217-13, Hayashi-machi, Takamatsu, Kagawa, 761-0301, Japan}

\author{Toshihiro Kawaguchi}
\affiliation{Department of Economics, Management and Information Science, Onomichi City University, 1600-2, Hisayamada, Onomichi, Hiroshima, 722-8506, Japan}

\author{Takashi Okamoto}
\affiliation{Department of Cosmosciences, Graduate School of Science, Hokkaido University, N10 W8, Kitaku, Sapporo, 060-0810, Japan}

\author{Masahiro Nagashima}
\affiliation{Faculty of Education, Bunkyo University, 3337, Minami-ogishima, Koshigaya, Saitama 343-8511, Japan}

\author{Taira Oogi}
\affiliation{Kavli Institute for the Physics and Mathematics of the universe, Todai Institutes for Advanced Study, the University of Tokyo, 5-1-5, Kashiwanoha, Kashiwa, 277-8583 Japan}

\correspondingauthor{Hikari Shirakata}
\email{shirakata@astro1.sci.hokudai.ac.jp}

\begin{abstract}
	We show the significance of the super-Eddington accretion for the cosmic growth of supermassive black holes (SMBHs)
	with a semi-analytical model for galaxy and black hole evolution.
	The model explains various observed properties of galaxies and
	active galactic nuclei at a wide redshift range.
	By tracing the growth history of individual SMBHs,
	we find that the fraction of the SMBH mass acquired during the super-Eddington accretion phases
	to the total SMBH mass becomes larger for less massive black holes and at higher redshift.
	Even at ${z \sim 0}$, SMBHs with ${> 10^9 M_\odot}$
	have acquired more than 50\% of their mass by super-Eddington accretions,
	which is apparently inconsistent with classical \Soltan argument.
	However, the mass-weighted radiation efficiency of SMBHs with ${> 10^8 M_\odot}$
	obtained with our model, is about 0.08 at ${z \sim 0}$,
	which is consistent with \Soltan argument within the observational uncertainties.
	We, therefore, conclude that \Soltan argument cannot reject the possibility that
SMBHs are grown mainly by super-Eddington accretions.
\end{abstract}


\section{Introduction}
\label{sec:intro}
	Almost all galaxies at ${z \sim 0}$ have a
	supermassive black hole (SMBH) at their center \citep[e.g.][]{Magorrian98}.
	SMBHs are considered to have grown by gas accretion \citep{Salpeter64,Lynden-Bell69}
	and BH-BH coalescence \citep[e.g.][]{KH00}.
	When the gas accretion occurs, the SMBH can be observed as
	an active galactic nucleus (AGN), which emits vast radiation
	when the material gets accreted onto the SMBH.
	The radiative energy per unit time, ${L_\mathrm{bol}}$
	(i.e. the bolometric luminosity of an AGN),
	can be described by the gas accretion rate, ${\dot{M}}$, as
	\begin{equation}
		L_\mathrm{bol} = \epsilon \dot{M} c^2,
	\end{equation}
	where $c$ and $\epsilon$ are the speed of light and the radiation efficiency, respectively.
	The mass increment per unit time for a black hole (BH), $\dot{M}_\mathrm{BH}$,
	is described as ${\dot{M}_\mathrm{BH} = (1-\epsilon) \dot{M}}$.

	As an indicator for how rapid an SMBH grows,
	the luminosity and accretion rate normalized by the Eddington limit
	have been employed.
	The Eddington luminosity and Eddington accretion rate are defined as
	\begin{align}
			&L_\mathrm{Edd} = \frac{4\pi c G m_p}{\sigma_\mathrm{T}} M_\mathrm{BH}, \\
			&\dot{M}_\mathrm{Edd} = L_\mathrm{Edd} / c^2,
		\end{align}
	where $G, m_p, \sigma_\mathrm{T}$, and $M_\mathrm{BH}$ are the gravitational constant, proton mass,
	cross-section for the Thomson scattering, and the mass of a BH, respectively.
	Assuming ${\epsilon \sim 0.1}$ for the sub-Eddington accretion rate,
	the gravitational force balances radiative pressure on the accreted gas
	in a spherical accretion and illumination case at
	${\dot{M} \sim 10 \dot{M}_\mathrm{Edd}}$ (i.e. ${L_\mathrm{bol} \sim L_\mathrm{Edd}}$).

	The radiation efficiency, $\epsilon$, depends on the SMBH spin,
	defined as ${a \equiv cJ/GM_\mathrm{BH}^2}$ ($J$ is the angular momentum of the BH),
	the Eddington ratio defined as ${\lambda_\mathrm{Edd} \equiv L_\mathrm{bol}/L_\mathrm{Edd}}$,
	and the innermost radius of the accretion disk.
	Assuming that the disk extends down to the innermost stable circular orbit,
	$\epsilon~$ is ${\sim 0.06}$ with ${a = 0}$ (i.e. the Schwarzschild BH) and
	${\sim 0.43}$ with ${a \to 1}$ (i.e. the Kerr BH) \citep{Bardeen70}.
	The properties of the accretion disk also depend on the Eddington ratio.
	The efficiency becomes maximum at ${\lambda_\mathrm{Edd} \sim 0.01}$ -- $1$
	and decreases at lower and higher $\lambda_\mathrm{Edd}$ regime
	\citep{Abramowicz88}
	due to the effect of the photon trapping (at ${\lambda_\mathrm{Edd} \gtrsim 1}$) and advection cooling
	\citep{Begelman78}.
	The dependence of the radiation efficiency on the Eddington ratio has been investigated
	by several authors \citep[e.g.][]{MK00,Watarai00,Kawaguchi03}.

	The contribution of the super-Eddington accretion to the cosmic growth of SMBHs is also important
	for understanding the co-evolution of SMBHs and galaxies via outflow \citep{Zamanov02, AKO05, Komossa08}
	and for constraining the mass of seed black holes.
	Observations have found luminous quasars at ${z > 6}$, whose SMBH masses are estimated as ${> 10^{9} M_\odot}$
	\citep[e.g.][]{Mortlock11,Wu15,Banados18}.
	Such SMBHs at high redshift tend to have shorter timescales from their birth to the observed time than local SMBHs.
	To explain the existence of such SMBHs at ${z \sim 6}$,
	SMBHs should have grown with a higher Eddington ratio or should form in very early epoch of the Universe,
	or their seed black hole mass should be large (namely, ${> 10^5 M_\odot}$).
	As a case study with a semi-analytic model of galaxy formation (hearafter SA model),
	\cite{PVS16} suggest that a luminous QSO at $z \sim 6$, SDSS J1148+5251,
	obtains $\sim 80$ \% of their mass at super-Eddington accretion rate
	owing to its dense and gas-rich environment.
	Also, various theoretical studies investigate the environment of the seed BHs of $z \sim 6$ QSOs
	with analytical methods \citep[e.g.][]{MR01,OSH08,TH09},
	hydrodynamical simulations \citep[e.g.][]{Hirano14,Chon16,Regan19}, and SA models \citep[e.g.][]{Valiante16J}.
	The environment and the effect of high radiative pressure of BHs
	with super-Eddington accretions have been investigated \citep[e.g.][]{Inayoshi16}.

	In our previous paper \citep{Shirakata19b}, we have presented the theoretical predictions
	of Eddington ratio distribution functions (ERDFs) of AGNs by using
	an SA model ``\textit{New Numerical Galaxy Catalogue ($\nu^2 GC$)}''
	\citep{nu2gc, Shirakata19}.
	In \cite{Shirakata19b}, we have found that SMBH growths
	via super-Eddington accretions become more significant at higher redshift and for less massive SMBHs.
	In this paper, we analyze our model data in the same way as \Soltan argument
	and conclude that our results of the significance of the super-Eddington accretion
	is consistent with \Soltan argument.
	Before showing our main results, we review \Soltan argument (Sec.\ \ref{sec:review})
	and briefly describe the growth model of SMBHs and analysis method (Sec.\ \ref{sec:model}).
	In Sec.\ \ref{sec:result},
	we show how significant the super-Eddington accretion is,
	by addressing the mass fraction of SMBHs acquired through super-Eddington growth, and
	compare the model results with observational data in the same manner as \Soltan argument.
	Finally, we discuss the consistency between our model results and \Soltan argument
	and summarize our results in Sec.\ \ref{sec:discussion}.
	Unless otherwise stated, we employ the $N$-body simulation \citep{Ishiyama15} with the box size $560 h^{-1}$ Mpc
	and $4096^3$ particles (the smallest halo mass is ${8.79 \times 10^9 M_\odot}$ with 40 dark matter particles).
	The parameters used in this paper are same as those in \cite{Shirakata19} and \cite{Shirakata19b}.

\section{Points to Review in \Soltan argument}
\label{sec:review}
	\Soltan argument \citep{Soltan82}
	is one of the well-known discussions on the cosmic growth of SMBHs.
	To understand the significance of the gas accretion for the cosmic growth of SMBHs,
	the following two values have been compared:
	\begin{align}
		&\rho_\mathrm{BH} (z = 0) = \int_{\log (M_\mathrm{BH,min})}^{\infty} M_\mathrm{BH} \Phi_\mathrm{BH} (M_\mathrm{BH}, z = 0) d \log M_\mathrm{BH}, \label{eq:massdens} \\
		&\rho_\mathrm{AGN}^\mathrm{acc} (z = 0) = \int_{\log (L_\mathrm{bol,min})}^{\infty}  d\log L_\mathrm{bol} \nonumber \\
		&\int_0^{\infty} \frac{(1-\epsilon) L_\mathrm{bol}}{\epsilon c^2}\Phi_\mathrm{AGN} (L, z)\frac{dt}{dz}dz \label{eq:lumdens},
	\end{align}
	where $\rho_\mathrm{BH} (z)$ and $\rho_\mathrm{AGN}^\mathrm{acc} (z)$ are the SMBH mass density
	at redshift $z$, and the accreted gas mass density from ${z = \infty}$ to $z$, respectively.
	The AGN luminosity function (LF), $\Phi_\mathrm{AGN}$, and SMBH mass function (MF), $\Phi_\mathrm{BH}$,
	should be observables.
	\citet[hereafter YT02]{YT02} compared $\rho_\mathrm{BH}$ and $\rho^\mathrm{acc}_\mathrm{AGN}$,
	which are obtained from type-1 QSO LFs, and found that they become comparable to each other
	when ${\epsilon \sim 0.1}$ -- $0.3$ is assumed.
	Therefore, SMBHs are considered to have grown mainly by the gas accretion, not BH-BH coalescence.

	\Soltan argument also constrains how rapid the SMBH growth is.
	Since ${\epsilon \sim 0.1}$ -- $0.3$ is consistent with the standard accretion disk \citep{SS73},
	it is often interpreted that SMBHs would have grown mainly by sub-Eddington accretions.
	This scenario is also supported by observational studies of the ERDF at ${z \sim 0}$
	\citep[e.g.][]{SW10}.
	One might conclude that the super-Eddington accretion is rare,
	and it is unimportant for the cosmic growth of the SMBHs.

However, other observational studies \citep[e.g.][]{MD04, Nobuta12, KS13} suggest
	that the super-Eddington accretion becomes more common at higher redshift.
	Also, several authors \citep[e.g.][]{Mortlock11,Wu15,Banados18} have found
	that QSOs (i.e. the brightest class of AGNs) at ${z \gtrsim 6}$ with ${M_\mathrm{BH} > 10^9 M_\odot}$
	are growing at ${\lambda_\mathrm{Edd} \gtrsim 1}$.
	On the theoretical side,
	some studies have found that the super-Eddington accretions
	should play a role in the cosmic growth of SMBHs
	by using hydrodynamic simulations \citep[e.g.][]{Angles-Alcazar17Nov} and
	SA models (e.g. \citealt{Shirakata19b}).
	These recent findings may conflict with \Soltan argument.

	One can argue that $\epsilon$, based on various observations,
	does not necessarily indicate that its value lie tightly between $0.1$ and $0.3$
	and that super-Eddington accretion could be the dominant mode for the SMBH growth as follows.
	First, as shown in \cite{Kawaguchi04June} in detail,
	${\rho_\mathrm{BH} (z = 0)}$ and ${\rho_\mathrm{AGN}^\mathrm{acc} (z = 0)}$,
	described in Eqs.\ \ref{eq:massdens} and \ref{eq:lumdens}, are governed by
	SMBHs with ${\sim 10^{8-9} M_\odot}$.
	As shown below, properties and amounts of BHs with ${< 10^8 M_\odot}$ are neglected.
	LFs and MFs are fitted by double power law functions.
	Each function has a steep slope at the luminous/massive end and a flat slope at the other end.
	Such a function has a point (hereafter ``knee'') at which the slope becomes $-1$.
	Assuming such functions, the integration values of LFs and MFs are mostly determined
	by the values around the knees,
	with a weak dependence on the slopes of the LFs/MFs below the knees.
	In \Soltan argument, ${\rho_\mathrm{BH} (z = 0)}$ is obtained by
	the SMBH MFs, whose knee is located at ${M_\mathrm{BH} \sim 1.4-3.5 \times 10^{8} M_\odot}$
	\citep{Shankar04}.
	Therefore, details (census and activity) on the SMBHs with ${M_\mathrm{BH} < 10^8 M_\odot}$
	has little influence on the ${\rho_\mathrm{AGN}^\mathrm{acc} (z = 0)}$ and ${\rho_\mathrm{BH} (z = 0)}$.

	Second, ${\rho_\mathrm{BH} (z = 0)}$ and $\rho^\mathrm{acc}_\mathrm{AGN}$ still have large uncertainties
	to put a constraint on the value of $\epsilon$ (see \cite{Novak13}, for more details).
	As for ${\rho_\mathrm{BH} (z = 0)}$, several empirical scaling relations between
	the SMBH mass and host bulge properties such as the luminosity, velocity dispersion, and
	stellar mass have been employed for obtaining the local SMBH MF.
	\citetalias{YT02} adopted ${\rho_\mathrm{BH} (z = 0) = (2.5 \pm 0.4) \times 10^5 M_\odot \mathrm{Mpc}^{-3}}$,
	assuming a relation between the SMBH mass and velocity dispersion.
	\cite{Vika09} derived ${\rho_\mathrm{BH} (z = 0) \sim 4.9 \times 10^5 M_\odot \mathrm{Mpc}^{-3}}$,
	using a relation between the SMBH mass and bulge luminosity.
	\cite{TV17} assumed an analytic expression for the local SMBH MF with a Schechter shape and a Gaussian scatter.
	They suggested that ${\rho_\mathrm{BH} (z = 0)}$ is ${4.3 (6.6) \times 10^5 M_\odot \mathrm{Mpc}^{-3}}$
	for a Gaussian scatter of 0.3 (0.5) dex.
	Considering the relation between SMBH mass and the S\'{e}rsic index for bulge surface density profile
	and its error, \cite{MSD16} estimated ${\rho_\mathrm{BH} (z = 0) = 2.04^{+1.16}_{-0.75} \times 10^5 M_\odot \mathrm{Mpc}^{-3}}$.
	Besides, the value of ${\rho_\mathrm{BH} (z = 0)}$ suffers another uncertainty
	if $M_\mathrm{BH}$ inferred from the emission line width is underestimated
	\citep{KH13_review}.
	As for ${\rho^\mathrm{acc}_\mathrm{AGN} (z = 0)}$,
	the bolometric correction from the AGN $B$-band or $X$-ray luminosity has large uncertainties.
	For example, the bolometric correction from the $B$-band luminosity, $C_B$, adopted in \citetalias{YT02} is $11.8$,
	while, according to a later study \citep{Marconi04}, the value is about $7$ for QSOs.
	Even if these uncertainties of the bolometric correction and ${\rho_\mathrm{BH} (z = 0)}$ are less than 1 dex,
	the value of $\epsilon$ could change drastically.

	The evolution of AGN LFs also involves large uncertainties.
	In \cite{Soltan82} and \citetalias{YT02}, ${\rho^\mathrm{acc}_\mathrm{AGN}(z = 0)}$ was estimated
	from optical AGN LFs. This means that only the contribution of ``type-1'' QSOs was considered.
	The values are ${4.7 \times 10^4 (0.1/\epsilon) M_\odot \mathrm{Mpc}^{-3}}$ \citep{Soltan82} and
	${2.1 \times 10^5 (C_B/11.8) [0.1 (1-\epsilon)/\epsilon] M_\odot \mathrm{Mpc}^{-3}}$ (\citetalias{YT02}).
	The evolution of the shape of optical AGN LFs has been under discussion, which largely affects the value of
	${\rho^\mathrm{acc}_\mathrm{AGN} (z = 0)}$.
	Optical AGN LFs have been assumed as a double power law shape:
	\begin{equation*}
		\Phi_\mathrm{AGN} (M_B,z) = \frac{\Phi^{*}_\mathrm{AGN}}{10^{0.4 \beta_1 [M_B-M^*_B(z)]} + 10^{0.4 \beta_2 [M_B-M^*_B(z)]}},
	\end{equation*}
	where $M_B$ is the $B$-band magnitude, ${\Phi^{*}_\mathrm{AGN}, \beta_1, \beta_2, M^{*}_B (z)}$ are adjustable parameters.
	In \citetalias{YT02}, the characteristic magnitude, $M^*_B$, was assumed to evolve as:
	\begin{equation}
		M^*_B(z) = -21.14 + 5\log h -2.5(1.36 z -0.27 z^2),
		\label{eq:MB}
	\end{equation}
	although recent observations find no such a strong evolution of $M^*_B$ especially at ${z > 3}$ \citep[e.g.][]{Akiyama18}.
	If there is no strong evolution of $M^*_B$, ${\rho^\mathrm{acc}_\mathrm{AGN} (z = 0)}$ obtained with the evolution of $M^*_B$
	(Eq.\ \ref{eq:MB}) is overestimated.
	Recent analysis, on the other hand, obtains ${\rho^\mathrm{acc}_\mathrm{AGN} (z = 0)}$ from
	the integration of $X$-ray AGN LFs, that is, the contribution of both type-1 and type-2 objects is considered.
	\cite{SWM09} found ${\rho^\mathrm{acc}_\mathrm{AGN} (z = 0)}$ should be ${\sim 5 \times 10^5 M_\odot \mathrm{Mpc}^{-3}}$
	with ${\epsilon \sim 0.075}$ so that ${\rho^\mathrm{acc}_\mathrm{AGN} (z = 0)}$ becomes ${\sim \rho_\mathrm{BH} (z = 0)}$.

	Given various uncertainties above, the dominance of the sub-Eddington accretion
	suggested by \Soltan argument is worth reassessing.
	In other words, the predominance of the super-Eddington accretion should be carefully considered.

\section{Methods}
	\label{sec:model}
	\subsection{SMBH growth in the semi-analytic model}
	\label{subsec:GrowthModel}
	We briefly describe the modeling of the SMBH growth
	\citep[see][for more details]{Shirakata19}.
	The seed BH mass is ${10^3 M_\odot}$ for all seed BHs
	\footnote{The seed BH mass does not largely affect the properties of AGNs and SMBHs
	at ${z \lesssim 6}$, since the seed mass is negligible compared with
	the total amount of the accreted gas onto a BH (see \citealt{Shirakata16}).}
	which are placed when galaxies newly form.
	We assume that an SMBH grows with its host bulge via starbursts induced by galaxy mergers and/or disk instabilities.
	The accreted gas mass onto the SMBH, $\Delta M_\mathrm{acc}$,
	and stellar mass formed by a starburst, $\Delta M_\mathrm{star,burst}$, have the following relation:
	\begin{equation}
		\Delta M_\mathrm{acc} = f_\mathrm{BH} \Delta M_\mathrm{star,burst},
	\end{equation}
	where ${f_\mathrm{BH} = 0.02}$ is chosen to reproduce the local BH mass -- bulge mass relation.

	The gas accretion rate is described as follows:
	\begin{equation}
		\dot{M} (t) = \frac{\Delta M_\mathrm{acc}}{t_\mathrm{acc}} \exp\left(-\frac{t-t_\mathrm{start}}{t_\mathrm{acc}}\right), \label{eq:Mdot}
	\end{equation}
	where $t_\mathrm{start}$ and $t_\mathrm{acc}$ are the starting time
	of the accretion and the accretion timescale per one accretion event, respectively.
	\footnote{$\dot{M}$ was described as $\dot{M}_\mathrm{BH}$ in \cite{Shirakata19}.}
	We use the same model of the accretion timescale as \cite{Shirakata19},
	${t_\mathrm{acc} = \alpha_\mathrm{bulge} t_\mathrm{dyn,bulge} + t_\mathrm{loss}}$.
	The first term of the right-hand side is proportional to the dynamical time of their host bulges,
	$t_\mathrm{dyn,bulge}$, where the value of the free parameter, $\alpha_\mathrm{bulge}$, is $0.58$.
	The second term represents the timescale for the angular momentum loss in the ``gas reservoir'' (e.g. circumnuclear disks)
	and accretion disk. We define $t_\mathrm{loss}$ as
	${t_\mathrm{loss,0} (M_\mathrm{BH}/M_\odot)^{\gamma_\mathrm{BH}} (\Delta M_\mathrm{acc}/M_\odot)^{\gamma_\mathrm{gas}}}$,
	where the values of the free parameters, $t_\mathrm{loss,0}$, $\gamma_\mathrm{BH}$, and $\gamma_\mathrm{gas}$, are
	1.0 Gyr, 3.5, and -4.0, respectively \citep[see][]{Shirakata19}.

	The AGN bolometric luminosity is described with ${\dot{m} \equiv \dot{M}/\dot{M}_\mathrm{Edd}}$
	(with ${\dot{M}_\mathrm{Edd} = L_\mathrm{Edd}/c^2}$) as
	\begin{equation}
		\lambda_\mathrm{Edd} = \frac{L_\mathrm{bol}}{L_\mathrm{Edd}} = \left[\frac{1}{1+3.5\{1+\tanh(\log(\dot{m}/\dot{m}_\mathrm{crit}))\}} + \frac{\dot{m}_\mathrm{crit}}{\dot{m}}\right]^{-1}.
		\label{eq:Lbol}
	\end{equation}
	We employ the formula, Eq.\ \ref{eq:Lbol}, based on \cite{Kawaguchi03},
	which takes into account various corrections
	(e.g. gravitational redshift, transverse Doppler effect).
	In this paper, ${\dot{m}_\mathrm{crit} = 10}$ is assumed.
	The radiation efficiency, $\epsilon$ (${= L_\mathrm{bol} / \dot{M} c^2}$),
	is defined as $\lambda_\mathrm{Edd} / \dot{m}$.
	In the super-Eddington regime (i.e. ${\dot{m} > \dot{m}_\mathrm{crit}}$),
	$\epsilon$ gradually decreases from $0.1$ at ${\dot{m} = \dot{m}_\mathrm{crit}}$.
	The value of $\epsilon$ is ${\sim 0.1}$ until ${\dot{m} \sim 30}$,
	and ${\sim 0.01}$ at ${\dot{m} \sim 600}$ (See Fig.\ 1 in \citealt{Shirakata19b}).
	For obtaining $B$-band luminosity of AGNs, we employ the bolometric correction
	obtained by \cite{Marconi04}.

	\subsection{Analysis for the significance of the super-Eddington growth}
	\label{subsec:Analythis}
	To evaluate the significance of the super-Eddington accretion,
	we calculate the accreted masses, $\Delta M_\mathrm{se} (z)$,
	$\Delta M_\mathrm{QSO} (z)$, and $\Delta M_\mathrm{se, QSO} (z)$,
	as follows:
	\begin{align}
		\begin{aligned}
			\Delta M_\mathrm{se} (z) = \sum\limits_{i} \int_{z}^{\infty} \dot{M}_{\mathrm{BH}, i} (z^\prime) \frac{dt}{dz^\prime} dz^\prime, \\
			i~\mathrm{for}~\dot{m} > \dot{m}_\mathrm{crit}, &\\
		\end{aligned}\\
		\begin{aligned}
			\Delta M_\mathrm{QSO} (z) = \sum\limits_{i} \int_{z}^{\infty} \dot{M}_{\mathrm{BH}, i} (z^\prime)  \frac{dt}{dz^\prime} dz^\prime, \\
			i~\mathrm{for}~L_\mathrm{bol} > 1.44 [10^{12} L_\odot], & \\
		\end{aligned}\\
		\begin{aligned}
			\Delta M_\mathrm{se, QSO} (z) = \sum\limits_{i} \int_{z}^{\infty} \dot{M}_{\mathrm{BH}, i} (z^\prime)  \frac{dt}{dz^\prime} dz^\prime, \\
			i~\mathrm{for}~\dot{m} > \dot{m}_\mathrm{crit} \, \, \mathrm{and}\, \, L_\mathrm{bol} > 1.44~[10^{12} L_\odot], &
		\end{aligned}
	\end{align}
	respectively.
	The summation ($i$) is taken for the SMBH subsample
	determined by the SMBH mass at a redshift  $z^\prime$, $M_\mathrm{BH} (z^\prime)$.
	We define QSOs as AGNs with ${L_\mathrm{bol} > 1.44~[10^{12} \, L_\odot]}$,
	which corresponds to the absolute $B$-band magnitude, $M_B$, ${\sim -23.7}$.
	By taking the ratio between $\Delta M_\mathrm{se}$ (or $\Delta M_\mathrm{QSO}$ or $\Delta M_\mathrm{se,QSO}$)
	and the sum of the total SMBH mass, ${\sum\limits_{i} M_{\mathrm{BH},i} (z)}$, we can estimate the importance
	of super-Eddington accretions (or of QSO phases or of super-Eddington accretions in QSO phases).

	We also estimate the mass-weighted mean radiation efficiency, $\bar{\epsilon} (z)$, as
	\begin{equation}
		\bar{\epsilon} (z) = \frac{\sum\limits_{i}\int_{z}^{\infty} \epsilon_i \dot{M}_{\mathrm{BH},i} (z^\prime)  \frac{dt}{dz^\prime} dz^\prime}{\sum\limits_{i} M_{\mathrm{BH},i} (z)},
		\label{eq:eps}
	\end{equation}
	where $\epsilon_i$ is the radiation efficiency obtained from $\lambda_\mathrm{Edd} / \dot{m}$ of the $i$-th SMBH.

	\section{Results}
	\label{sec:result}
	\subsection{Comparison with \Soltan argument}
	\label{subsec:Soltan}
	We make a straightforward comparison with \Soltan argument.
	Since our model reproduces observed AGN LFs at ${0 < z < 6}$ and local SMBH MF
	\citep{Shirakata19},
	the model should return the consistent result with \Soltan argument,
	i.e. ${\rho_\mathrm{AGN}^\mathrm{acc} (z = 0)}$ obtained from the QSO luminosity functions
	(with ${\epsilon \sim 0.1}$ -- $0.3$) becomes ${\sim \rho_\mathrm{BH} (z = 0)}$.

	First, we obtain ${\rho_\mathrm{AGN}^\mathrm{acc} (z = 0)}$ from the QSO luminosity functions
	at ${0 \lesssim z \lesssim 5}$ by our model.
	Following the conventional procedure, we fix ${\epsilon = 0.1}$
	in converting the AGN luminosity into the mass accretion rate,
	while we have calculated the AGN luminosity in the model from Eqs.\ \ref{eq:Mdot} and \ref{eq:Lbol}.
	The resultant ${\rho_\mathrm{AGN}^\mathrm{acc} (z = 0)}$ is ${3.6 \times 10^5 M_\odot \mathrm{Mpc}^{-3}}$,
	including all type-1 and type-2 AGNs with ${M_B < -23}$, which is only $1.7$ times larger than
	the ${\rho_\mathrm{AGN}^\mathrm{acc} (z = 0)}$ only for type-1 AGNs obtained by \citetalias{YT02}
	since the evolution of QSO LFs assumed in \citetalias{YT02} seems to be
	inconsistent with recent observational results (see also Sec.\ \ref{sec:review}).
	The value of ${\rho_\mathrm{AGN}^\mathrm{acc} (z = 0)}$ obtained by our model
	becomes ${9.4 \times 10^4 M_\odot \mathrm{Mpc}^{-3}}$,
	assuming that type-1 QSOs account for $16$ \% of total AGNs with ${M_B < -23}$ \citep{Shirakata19}.
	\footnote{In \cite{Shirakata19}, the fraction of type-1 QSOs depends
	both on redshift and AGN luminosity so that we obtain the same bolometric AGN LF
	from the AGN $X$-ray (2-10 keV) and UV (1450 \AA).
	Since the dependencies on redshift and luminosity are weak, we employ a constant value (16 \%).}

	Second, we determine the lower limit of the integration, $M_\mathrm{BH,min}$, in Eq.\ \ref{eq:massdens}.
	By integrating the model SMBH MF at ${z \sim 0}$ obtained by our model,
	we find that ${\rho_\mathrm{AGN}^\mathrm{acc} (z = 0)}$ becomes ${\sim \rho_\mathrm{BH} (z = 0)}$,
	when ${M_\mathrm{BH,min}}$ is ${\sim 1.7 \times 10^8 M_\odot}$ (under the assumption that
	all AGNs are type-1) and ${\sim 1.1 \times 10^9 M_\odot}$
	(under the assumption that type-1 QSOs account for $16$ \% of total AGNs with ${M_B < -23}$, \citealt{Shirakata19}).
	Given that the knee of the SMBH MF at ${z \sim 0}$ places at ${1.4-3.5 \times 10^{8} M_\odot}$
	\citep{Shankar04}, the model result without obscuration (i.e. the same assumption as \citetalias{YT02})
	is consistent with \Soltan argument since the value of $\rho_\mathrm{BH}$ is determined
	by the value around the knee, as described in Sec.\ \ref{sec:review}.
	Therefore, we conclude that the SMBHs with ${M_\mathrm{BH} \gtrsim 1.7 \times 10^8 M_\odot}$ have
	grown mainly by gas accretions during QSO phases.

	\subsection{Growth history of individual SMBHs}
	\label{subsec:theory}
\begin{figure}
			\begin{center}
				\includegraphics[width=\hsize]{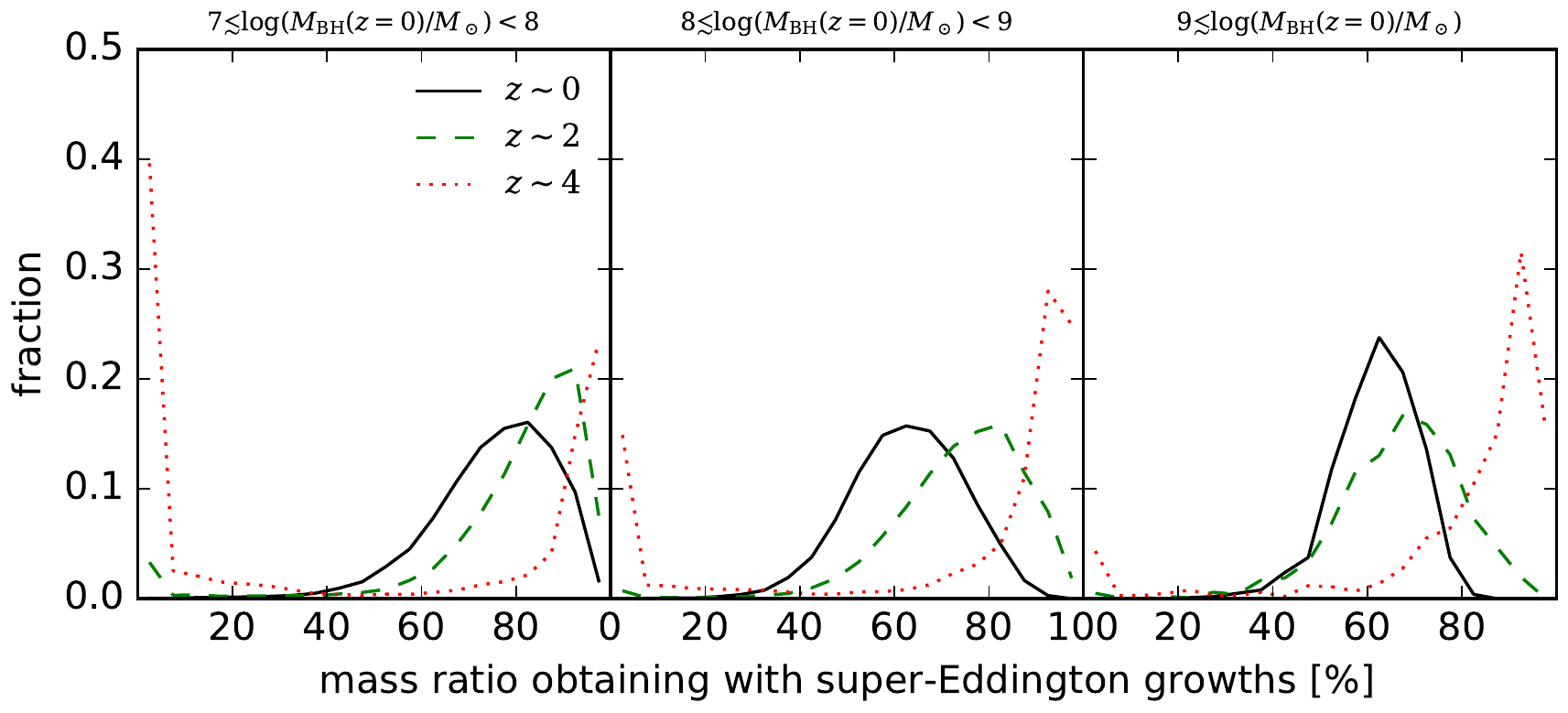}
			\end{center}
			\caption{The distribution of $M_{\mathrm{se},i} / M_{\mathrm{BH},i}$
			with different SMBH mass bins (${\log(M_\mathrm{BH} (z = 0) /M_\odot) = [7,8], [8,9]}$, and $> 9$, from left to right panels)
			and different redshift (${z \sim 0, 2}$, and $4$, as shown in black solid, green dashed, and red dotted lines, respectively).}
			\label{fig:evolve}
	\end{figure}

	In order to show the redshift evolution in another way,
	we trace the evolution of individual SMBHs with ${\log(M_\mathrm{BH} (z^\prime=0)/M_\odot) = [7,8], [8,9],}$ and $> 9$
	in Fig.\ \ref{fig:evolve}.
	Each panel shows the distribution of ${M_{\mathrm{se},i} / M_{\mathrm{BH},i}}$
	at ${z  = 0, 2,}$ and $4$ for each SMBH mass bin.
	By investigating the distribution, we assess how typical AGNs at each redshift- and mass-bins behave.
	The peak of the distribution moves toward the higher $M_{\mathrm{se},i} / M_{\mathrm{BH},i}$
	at higher redshift, meaning that higher-$z$ SMBHs have greater contribution of super-Eddington accretion.
	However, even SMBHs with ${M_\mathrm{BH} (z = 0) > 10^9 M_\odot}$,
	more than 50 \% of SMBHs acquire $> 60$ \% of their mass by super-Eddington accretions.
	The same suggestion is obtained in \cite{Shirakata19b}.

\begin{figure}
			\begin{center}
				\includegraphics[width=\hsize]{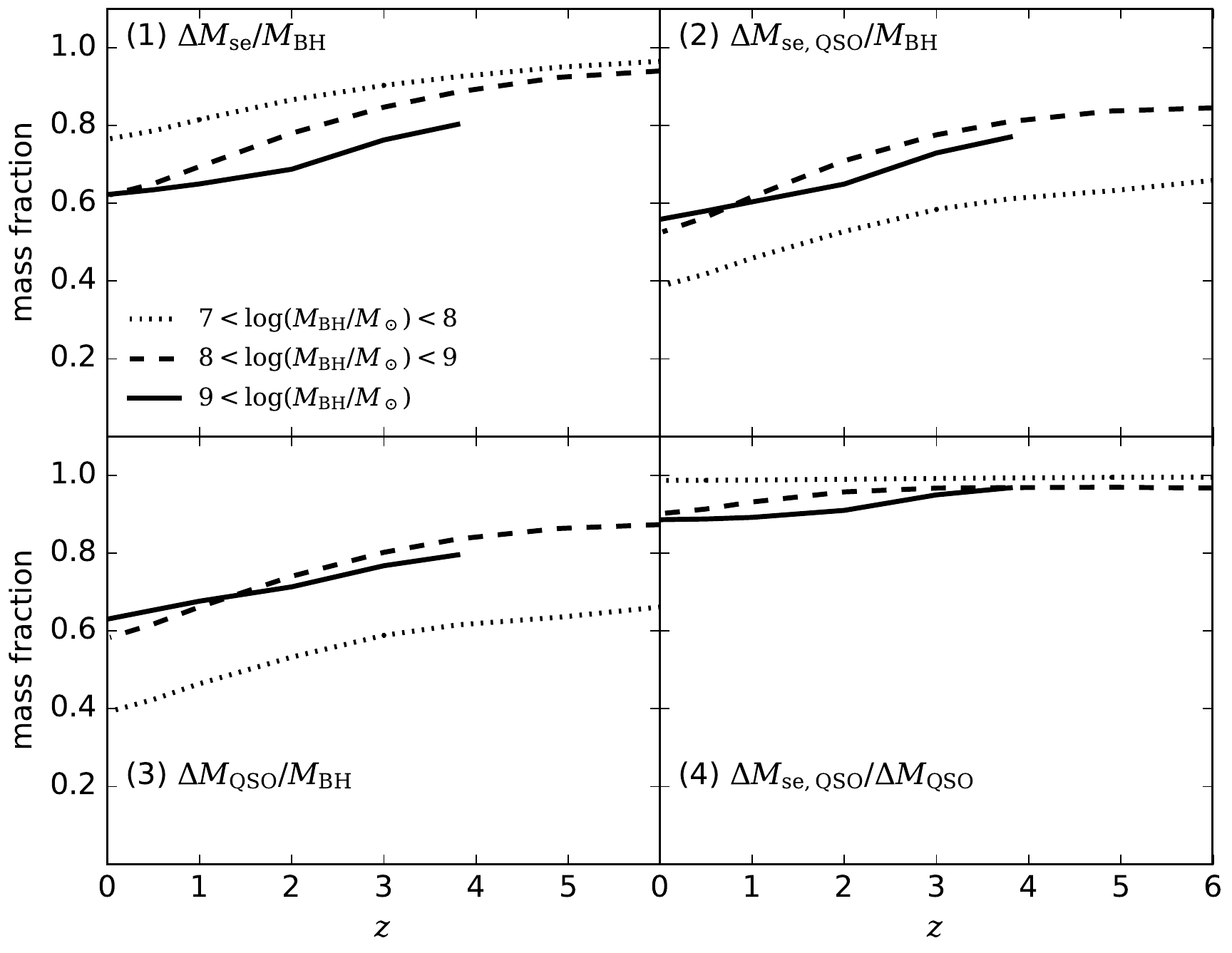}
			\end{center}
			\caption{The values of the four types of mass fraction: $\Delta M_\mathrm{se} (z)/\sum\limits_{i} M_{\mathrm{BH},i} (z)$ (top left),
			$\Delta M_\mathrm{se, QSO} (z)/\sum\limits_{i} M_{\mathrm{BH},i} (z)$ (top right),
			$\Delta M_\mathrm{QSO} (z)/\sum\limits_{i} M_{\mathrm{BH},i} (z)$ (bottom left), and
			$\Delta M_\mathrm{se, QSO} (z)/\Delta M_\mathrm{QSO} (z)$ (bottom right).
		Black dotted, dashed, solid lines are results of SMBHs with ${\log(M_\mathrm{BH} (z = 0) /M_\odot) = [7,8], [8,9],}$ and $> 9$, respectively.}
			\label{fig:mass}
	\end{figure}

	Fig.\ \ref{fig:mass} shows the evolution of
	$\Delta M_\mathrm{se} (z)/\sum\limits_{i} M_{\mathrm{BH},i} (z)$,
	$\Delta M_\mathrm{se, QSO} (z)/\sum\limits_{i} M_{\mathrm{BH},i} (z)$,
	$\Delta M_\mathrm{QSO} (z)/\sum\limits_{i} M_{\mathrm{BH},i} (z)$, and
	$\Delta M_\mathrm{se, QSO} (z)/\Delta M_\mathrm{QSO} (z)$, respectively.
	The vertical axis in the first panel (top left panel) corresponds
	to horizontal axis of Fig.\ \ref{fig:evolve}.
	From this figure, we find that
	\begin{enumerate}
		\item for SMBHs with ${10^7 M_\odot < M_\mathrm{BH} (z = 0) < 10^8 M_\odot}$, about half of the super-Eddington growth
			occurs at less luminous AGN phases, not QSO phases (by comparing panels (1) and (2)),
		\item even SMBHs with ${10^9 M_\odot < M_\mathrm{BH} (z = 0)}$, SMBHs do not acquire their whole mass
			in QSO phases (the panel (3)), and
		\item typical QSOs at any redshift and mass bins have acquired
			their masses mostly at super-Eddington phases (the panel 4).
			In other words, a significant fraction of QSOs at any redshift are
			expected to show ${\lambda_\mathrm{Edd} \gtrsim 1}$.
			It is consistent with the result by \cite{Collin02} who found that
			about half of nearby bright QSOs (PG QSOs) are accreting close to or exceeding
			the Eddington rate.
	\end{enumerate}

\begin{figure}
			\begin{center}
				\includegraphics[width=\hsize]{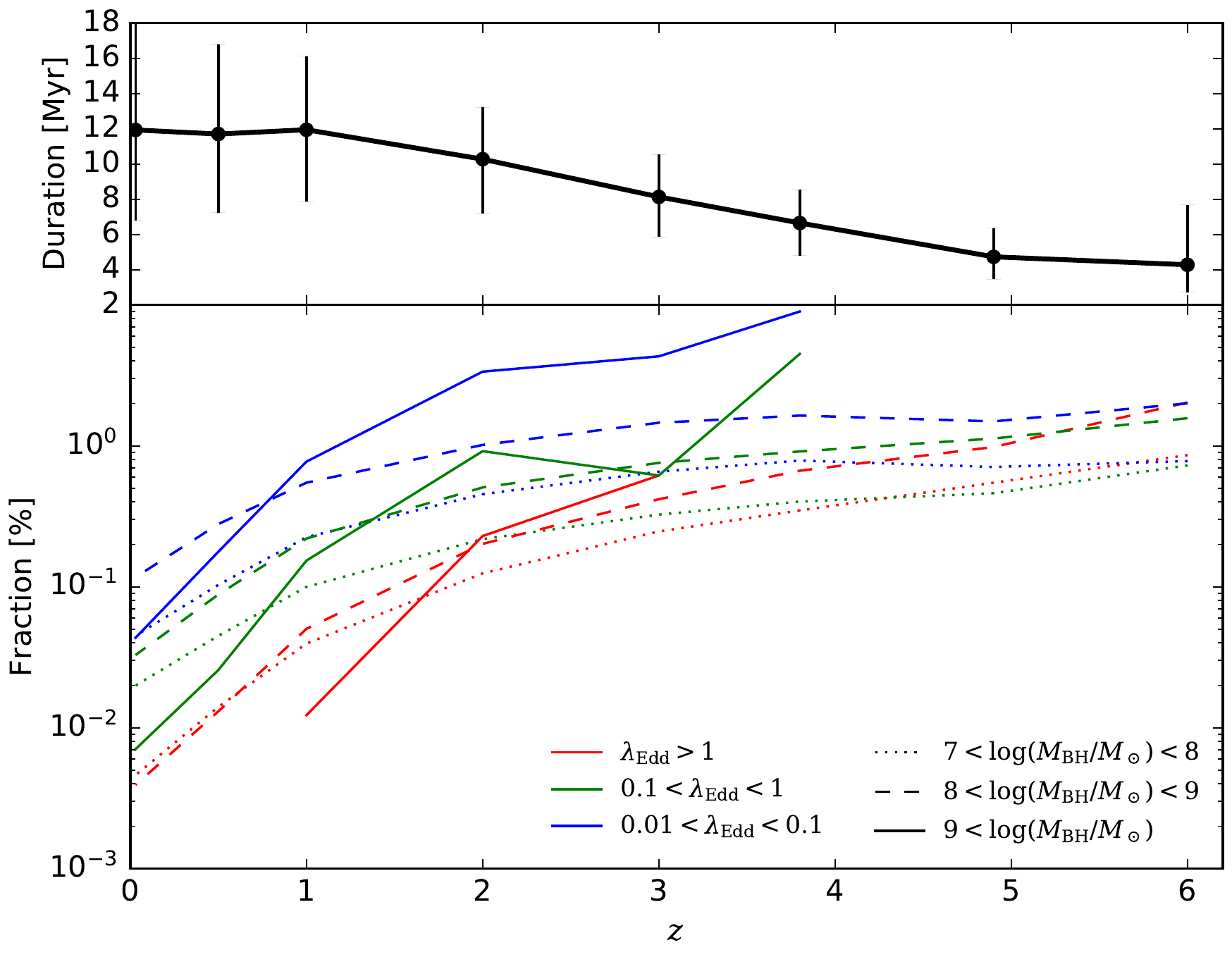}
			\end{center}
			\caption{\protect\textit{Top}: The median value of the duration
			of super-Eddington accretions. The error bar shows the $25-75$ percentile.
			\protect\textit{Bottom}: The fraction of active SMBHs with different $\lambda_\mathrm{Edd}$ range
			in total (i.e. active and non active) SMBHs at an output time in our model:
			$\lambda_\mathrm{Edd} = 0.01-0.1$ (blue), $0.1-1$ (green), and $> 1$ (red).
			The dotted, dashed, and solid lines show the different SMBH mass range
			(same as Fig.\ \protect\ref{fig:mass}).}
			\label{fig:duty}
	\end{figure}

\begin{figure}
			\begin{center}
				\includegraphics[width=\hsize]{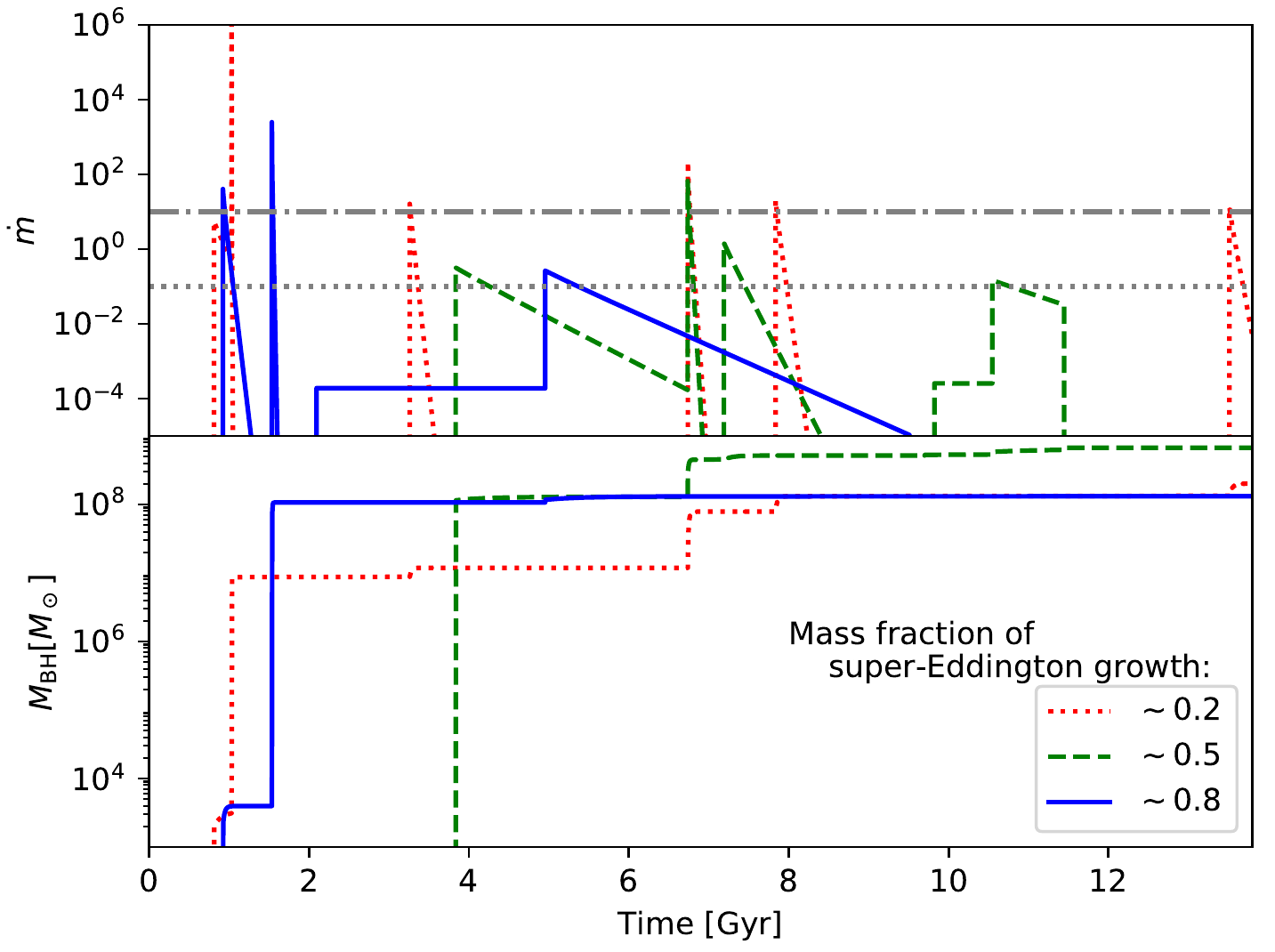}
			\end{center}
			\caption{The history of $\dot{m}$ for selected SMBHs with $M_\mathrm{BH} = 10^{8-9} M_\odot$
			at $z \sim 0$. They have acquired $\sim 20, 50,$ and $80$ \% of their mass with $\dot{m} > 10$ (grey dot-dashed line),
			namely, super-Eddington accretions (red dotted, green dashed, and blue solid lines, respectively).
			Since most QSOs and Seyfert galaxies are observed with $\lambda_\mathrm{Edd} > 0.01$
			(i.e. $\dot{m} > 0.1$), observable AGN phases place above the grey dotted line.
			The bottom panel shows the corresponding increment of $M_\mathrm{BH}$.}
			\label{fig:mdot_hist}
	\end{figure}

	Even super-Eddington growth plays a role in the cosmic growth of SMBHs,
	the probability with which we can observe super-Eddington accreting SMBHs is low.
	We investigate the duration of each super-Eddington accretion episode in Fig.\ \ref{fig:duty}.
	The median value of the duration obtained from our model
	becomes shorter at higher redshift; $12$ Myr at $z \sim 0$ and $4$ Myr at $z \sim 6$,
	although super-Eddington accretion becomes more common at higher redshift.
	The decreasing trend with redshift results from shorter $t_\mathrm{acc}$ at higher redshift.
	Due to the short duration of the super-Eddington phase, the fraction of super-Eddington accreting SMBHs
	with $10^{7-8} M_\odot$ at an output time in our model, for example,
	is only $\sim 4 \times 10^{-3}$ \% among all SMBHs (and $\sim 6.6$ \% in all $\lambda_\mathrm{Edd} > 0.01$ AGNs) at $z \sim 0$,
	and $\sim 1$ \% (and $\sim 36.3$ \%) at $z \sim 6$ (the bottom panel of Fig.\ \ref{fig:duty}).
	Examples of actual $\dot{m}$ history of each SMBH is shown in Fig.\ \ref{fig:mdot_hist}.
	We choose three SMBHs with $M_\mathrm{BH} = 10^{8-9} M_\odot$ at $z \sim 0$, which have acquired
	$\sim 20, 50, $ and $80$ \% of their mass with super-Eddington accretions.
	As shown in Fig.\ \ref{fig:duty}, super-Eddington accretions do not last long time
	and accretion rates stay at $\dot{m} < 10$ for most of their lives.

	In the panel (1) of Fig.\ \ref{fig:mass}, we also show that the less massive SMBHs
	have the higher value of $\Delta M_\mathrm{se}/M_\mathrm{BH}$,
	which was also mentioned in detail in \cite{Shirakata19b} with the same model as this paper.
	Recent observations show that AGNs with the less massive BHs tend to have the higher
	values of $\lambda_\mathrm{Edd}$ at $z \sim 0$, which is qualitatively consistent with our model prediction.
	For example, observational samples of \cite{Dong12} and \cite{Liu18}
	with $10^{5-6.5} M_\odot$ of SMBH mass at $z \sim 0$
	have the $\lambda_\mathrm{Edd}$ distribution peaking at $\log(\lambda_\mathrm{Edd}) \sim -0.4$.
	For more massive SMBHs, Fig.\ 3 of \cite{SW10} shows the number distribution of $\lambda_\mathrm{Edd}$
	for AGNs with $M_\mathrm{BH} \sim 10^{6-9} M_\odot$ from the Hamburg/ESO Survey.
	The distribution peaks at $\log(\lambda_\mathrm{Edd}) \sim -1$,
	which is smaller than for the less massive SMBHs.
	The data of SDSS QSO shows the similar distribution at $z = 0.4$,
	peaking at $\log(\lambda_\mathrm{Edd}) \sim -0.8$, with $M_\mathrm{BH} \sim 10^{8} M_\odot$
	\citep{KS13}.

\section{Discussion and conclusions}
	\label{sec:discussion}
	We have investigated the growth history of SMBHs
	by using an SA model, which explains various observed properties
	of galaxies and AGNs at a wide redshift range.
	Since \Soltan argument is based on the AGN LFs
	at $z < 6$ and the local SMBH MF, the growth processes of SMBHs
	in our model should satisfy the constraints by the argument,
	which imply that the SMBHs have grown mainly via sub-Eddington accretion events.
	When we adopt the radiation efficiency of $\sim 0.1$ in the QSOs in our model AGNs
	and estimate the accreted mass during the QSO phase from $z = 6$ to $0$,
	we obtain the mass density of SMBHs with $M_\mathrm{BH} > 1.7 \times 10^8 M_\odot$
	at $z = 0$. This is in line with \Soltan argument.
	We, however, find that even SMBHs with $M_\mathrm{BH} > 10^9 M_\odot$ at $z \sim 0$
	have acquired more than 50 \% of their mass by the super-Eddington accretion
	and a significant fraction of QSOs at any redshift are expected to have undergone the super-Eddington accretion.

	We have noted that super-Eddington accretions are difficult to observe
	since the durations of each super-Eddington accretion event is short;
	$12$ Myr at $z \sim 0$ and $4$ Myr at $z \sim 6$.
	Due to the short duration of the super-Eddington phase,
	the fraction of super-Eddington accreting SMBHs at an output time in our model
	is only $\sim 4 \times 10^{-3}$ \% among all SMBHs at $z \sim 0$,
	and $\sim 1$ \% at $z \sim 6$ (Fig.\ \ref{fig:duty}).
	In other words, our model predicts that only $\sim 6.6$ \% of SMBHs
	in observed AGNs with $\lambda_\mathrm{Edd} > 0.01$ are
	$\lambda_\mathrm{Edd} > 1$ at $z \sim 0$.
	Therefore, just from observational data,
	we underestimate the importance of super-Eddington
	accretions for cosmic growth of SMBHs.

	One might think that if SMBHs acquires their mass mainly by super-Eddington accretions,
	then the SMBH mass at $z \sim 0$ cannot be provided by
	observed QSOs with sub-Eddington accretions.
	Our model predictions, however, show no contradiction
	with the observations on which \Soltan argument bases as we discussed above.

\begin{figure}
			\begin{center}
				\includegraphics[width=\hsize]{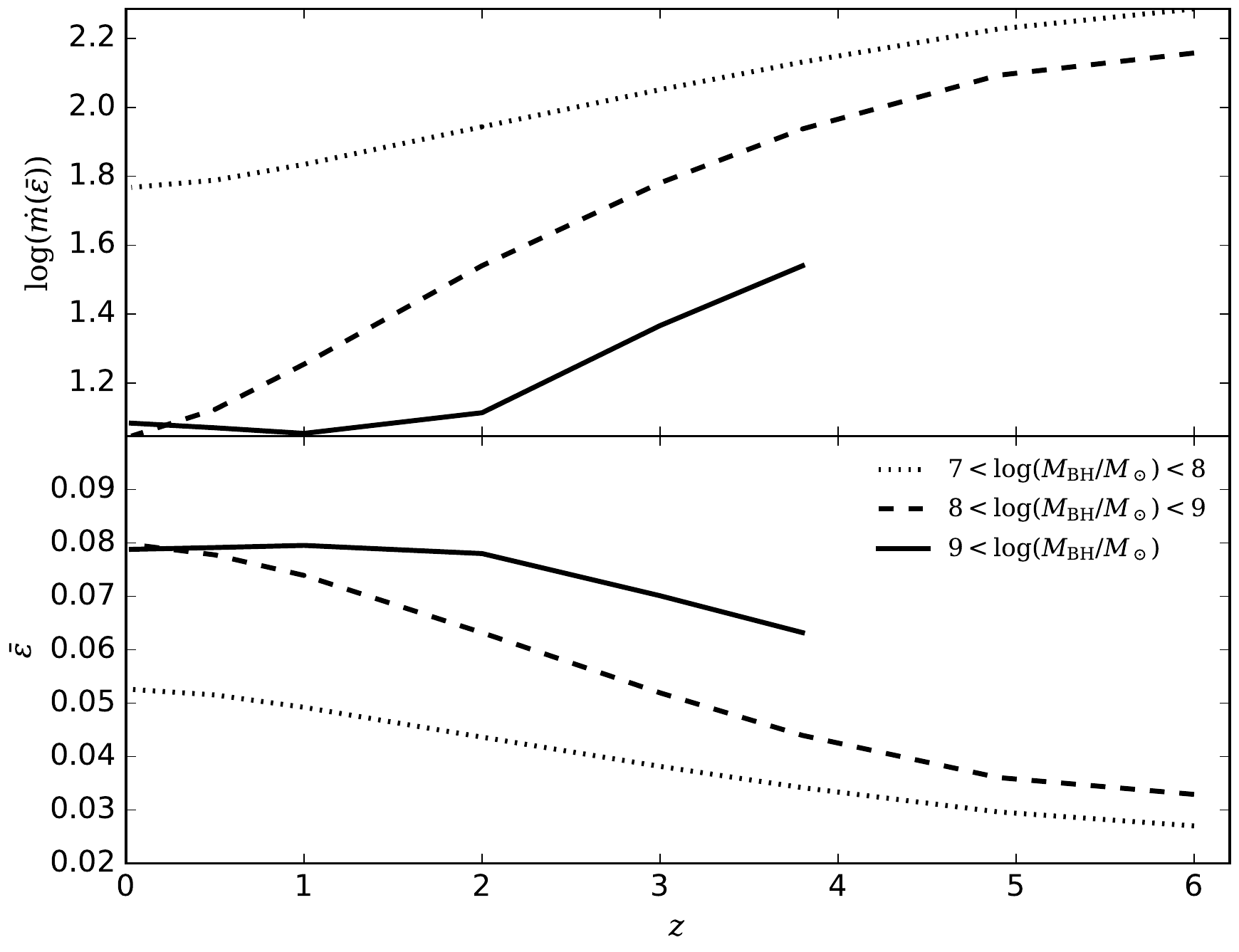}
			\end{center}
			\caption{The median value of $\dot{m}$ (top) obtained from the
			mass weighted radiation efficiency, $\bar{\epsilon}$ (bottom) using Eq.\ \protect\ref{eq:Lbol}.
			The lines are the same as Fig.\ \protect\ref{fig:mass}.}
			\label{fig:eta}
	\end{figure}

	To understand well these apparently contradictory results,
	we investigate the evolution of the mass weighted radiation efficiency, $\bar{\epsilon}$ (Eq.\ \ref{eq:eps}),
	for different SMBH mass bins in Fig.\ \ref{fig:eta}.
	Since super-Eddington accretion is more common at higher redshift
	(see also the top panel of Fig.\ \ref{fig:eta}),
	$\bar{\epsilon}$ is small at high redshift
	(e.g. $\sim 0.04$ at $z \sim 4$ with ${\log(M_\mathrm{BH} (z = 4)/M_\odot) = [8,9]}$).
	In contrast, $\bar{\epsilon}$ becomes larger, $\sim 0.08$, at $z \sim 0$ for SMBHs with
	${\log(M_\mathrm{BH} (z = 0)/M_\odot) > 8}$.
	This value of ${\bar{\epsilon} \sim 0.08}$ is not rejected by \Soltan argument given various uncertainties
	discussed in Sec.\ \ref{sec:review},
	although more than half of SMBHs with ${M_\mathrm{BH} (z = 0) > 10^8 M_\odot}$ in our model
	acquire their mass mainly by super-Eddington accretions as shown in Fig.\ \ref{fig:mass}.
	This is because, (1) $\epsilon$ decreases slowly towards higher $\dot{m}$,
	and (2) the accretion with higher $\dot{m}$ (${\dot{m} \gg \dot{m}_\mathrm{crit}}$)
	is rarer, considering the shape of ERDFs \citep{Shirakata19b}.

	Conclusions of this paper are as follows:
\begin{enumerate}
	\item{The Soltan argument does not reject the possibility that SMBHs are grown mainly by super-Eddington accretions,
		because assumptions employed in the classical Soltan argument (e.g., $z$-evolution of AGN LFs)
		do not match the current observational data.}
	\item{When we take current statistical data (estimated values and their uncertainties)
		of galaxies and AGNs at face value,
		our semi-analytical model suggests that SMBHs have grown mainly by super-Eddington accretions.}
	\item{Further observations with smaller uncertainties will judge the conclusion of this paper.}
\end{enumerate}

	As we discuss in Sec.\ \ref{sec:review}, uncertainties in observational
estimates for $\rho_\mathrm{BH}$ and $\rho^\mathrm{acc}_\mathrm{AGN}$
is crucial in \Soltan argument.
For example, relations between SMBH mass and properties of the host galaxies are important.
SMBHs also reside in the disk dominated galaxies. In such case, SMBH mass is difficult to estimate
from host galaxy's properties and the uncertainty of $\rho_\mathrm{BH}$ will become larger.
The difference of AGN SEDs among individual AGNs and SMBHs
and an increase of the number of AGNs with $M_\mathrm{BH} < 10^{7} M_\odot$ are also important.
In this paper, we employ the bolometric correction independently of the SMBH mass or Eddington ratio,
which is the same treatment as different semi-analytic models \citep[e.g.][]{Fanidakis12, Hirschmann12, Menci14}.
Theoretical and observational understandings of AGN SEDs will help to improve the discussion.
When the uncertainties of observational estimates for $\rho_\mathrm{BH}$ and $\rho^\mathrm{acc}_\mathrm{AGN}$
are reduced and/or when the shape of observed AGN LFs and SMBH MFs are determined with smaller errors,
our conclusion in this paper would be re-evaluated.
However, the main conclusion of this paper that SMBHs predominatly acquire their mass
throught super-Eddington accretion remains unchanged (see Appendix \ref{App:robust}).
Also, if the averaged radiation efficiency at higher redshift
is estimeted, it will be helpful to judge the importance of super-Eddington accretions
on the cosmic growth of SMBHs.
As an example, \cite{Davies19} estimate the averaged radiation efficiency of two
$z > 7$ SMBHs from their mass and ionized region size.
The estimated values are 0.08 and 0.1.
The increase of the sample size is needed for statistical discussions.

\acknowledgments
We appreciate the detailed review and useful suggestions
by the anonymous referee, which have improved our paper.
H. ~Shirakata has been supported by the Sasakawa Scientific Research Grant from
the Japan Science Society (29-214), JSPS KAKENHI (18J12081), and
a grant from the Hayakawa Satio Fund awarded by the Astronomical Society of Japan.
T. ~Kawaguchi has been supported by JSPS KAKENHI (17K05389)
and the grant from the Urakami Scholarship Foundation.
T.~Okamoto is financially supported by
MEXT KAKENHI Grant (18H04333) and the Grant-in-Aid (19H01931).
M.~Nagashima has been supported by the Grant-in-Aid (17H02867 and 18H05437)
from the MEXT of Japan.
This work was also supported in part by World Premier International Research Center Initiative (WPI),
MEXT, Japan, by MEXT Priority Issue 9 on Post-K Computer
(Elucidation of the Fundamental Laws and Evolution of the Universe), and by JICFuS.

\appendix

\section{The effect of the parameter choice}
\label{App:robust}
Here we show the robustness of our main result,
		the predominance of the super-Eddington accretion for the cosmic growth of SMBHs.
		Our model reproduces current observational AGN LFs and SMBH MFs.
		If observational AGN LFs and SMBH MFs largely changed their shapes by future observations,
		the model parameters should take different values
		and the model prediction about the dominance of the super-Eddington accretion might change.
		We present the results with different parameter choices below and show that
		the main conclusion of this paper is quite general and does not change
		even with other extreme parameter choices.

		As described in Sec. \ref{sec:model}, the gas accretion rate is modeled as
		$\dot{M} (t) = \frac{\Delta M_\mathrm{acc}}{t_\mathrm{acc}} \exp\left(-\frac{t-t_\mathrm{start}}{t_\mathrm{acc}}\right)$,
		where $\Delta M_\mathrm{acc}$ is the total accreted gas mass.
		In principle, the Eddington ratio for an SMBH with a given mass becomes smaller
		with the smaller value of $\Delta M_\mathrm{acc}$ or larger value of $t_\mathrm{acc}$.
		We note that if we make $t_\mathrm{acc}$ longer, bright AGNs become difficult to emerge
		since we assume that the maximum accretion rate is given as
		$\Delta M_\mathrm{acc} / t_\mathrm{acc}$.

\begin{figure}
			\begin{center}
				\includegraphics{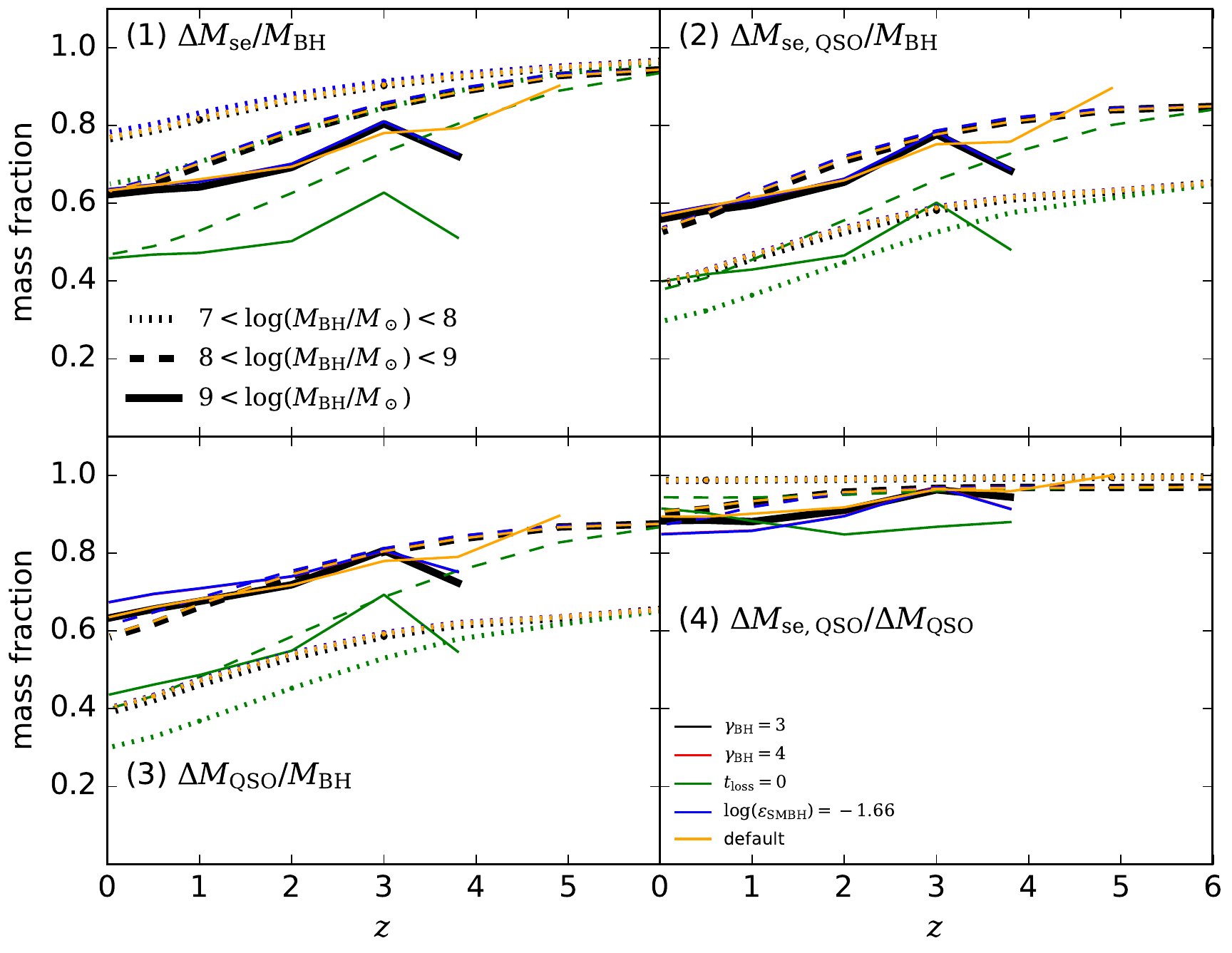}
			\end{center}
			\caption{The same figure as Fig.~\protect\ref{fig:mass} but outcomes of different
			combinations of parameters are included.
			The color shows the result with different parameter sets,
			that is, red, green, blue, and orange
			indicate $\gamma_\mathrm{BH} = 3$ and $\gamma_\mathrm{BH} = 4$,
			$t_\mathrm{loss} = 0$, and $\log(\epsilon_\mathrm{SMBH}) = -1.66$, respectively.
			The black lines indicate the default parameter set shown in Fig. ~\protect\ref{fig:mass}.}
			\label{fig:Eddington_dominant_app}
	\end{figure}

		We test several combinations of parameters, in which one parameter has a value
		different from the default one with the remaining parameters having the default ones.
		The parameter values we test here are $\gamma_\mathrm{BH} = 3$, $\gamma_\mathrm{BH} = 4$, $t_\mathrm{loss} = 0$,
		and $\log(\epsilon_\mathrm{SMBH}) = -1.66$ (corresponding to a stronger AGN feedback
		which quenches the formation of massive galaxies at $z < 1$; \citet{nu2gc} and \citet{Shirakata19}).
		When we choose $\gamma_\mathrm{BH} = 3, 4$ or $t_\mathrm{loss} = 0$, we cannot reproduce the shape
		of AGN LFs especially at $z < 1$.
		When we choose $\log(\epsilon_\mathrm{SMBH}) = -1.66$, massive galaxies at $z < 1$ cannot form
		because of the strong AGN feedback.
		The case with $\gamma_\mathrm{BH} = 4$, the accretion timescale of SMBHs becomes
		longer and possibly exceed the cosmic age (depending on the SMBH mass).
		Fig. \ref{fig:Eddington_dominant_app} shows the results with different combinations of parametes,
		which is the same figure as Fig. \ref{fig:mass} (the fraction of mass acquired through super-Eddington accretion).
		Even the most extreme case with $\gamma_\mathrm{BH} = 4$, $\Delta M_\mathrm{se} / \Delta M_\mathrm{BH}$ is reduced only $\sim 30$ \%.
		As shown here, drastic changes of free parameters related with the SMBH growth
		do not have large impacts on the main results of this work.

\bibliographystyle{aasjournal}
\bibliography{Astrophysics} 

\begin{thebibliography}{}
\expandafter\ifx\csname natexlab\endcsname\relax\def\natexlab#1{#1}\fi
\providecommand{\url}[1]{\href{#1}{#1}}
\providecommand{\dodoi}[1]{doi:~\href{http://doi.org/#1}{\nolinkurl{#1}}}
\providecommand{\doeprint}[1]{\href{http://ascl.net/#1}{\nolinkurl{http://ascl.net/#1}}}
\providecommand{\doarXiv}[1]{\href{https://arxiv.org/abs/#1}{\nolinkurl{https://arxiv.org/abs/#1}}}

\bibitem[{{Abramowicz} {et~al.}(1988){Abramowicz}, {Czerny}, {Lasota}, \&
  {Szuszkiewicz}}]{Abramowicz88}
{Abramowicz}, M.~A., {Czerny}, B., {Lasota}, J.~P., \& {Szuszkiewicz}, E. 1988,
  \apj, 332, 646, \dodoi{10.1086/166683}

\bibitem[{{Akiyama} {et~al.}(2018){Akiyama}, {He}, {Ikeda}, {Niida}, {Nagao},
  {Bosch}, {Coupon}, {Enoki}, {Imanishi}, {Kashikawa}, {Kawaguchi}, {Komiyama},
  {Lee}, {Matsuoka}, {Miyazaki}, {Nishizawa}, {Oguri}, {Ono}, {Onoue}, {Ouchi},
  {Schulze}, {Silverman}, {Tanaka}, {Tanaka}, {Terashima}, {Toba}, \&
  {Ueda}}]{Akiyama18}
{Akiyama}, M., {He}, W., {Ikeda}, H., {et~al.} 2018, \pasj, 70, S34,
  \dodoi{10.1093/pasj/psx091}

\bibitem[{{Angl{\'e}s-Alc{\'a}zar} {et~al.}(2017){Angl{\'e}s-Alc{\'a}zar},
  {Faucher-Gigu{\`e}re}, {Quataert}, {Hopkins}, {Feldmann}, {Torrey}, {Wetzel},
  \& {Kere{\v{s}}}}]{Angles-Alcazar17Nov}
{Angl{\'e}s-Alc{\'a}zar}, D., {Faucher-Gigu{\`e}re}, C.-A., {Quataert}, E.,
  {et~al.} 2017, \mnras, 472, L109, \dodoi{10.1093/mnrasl/slx161}

\bibitem[{{Aoki} {et~al.}(2005){Aoki}, {Kawaguchi}, \& {Ohta}}]{AKO05}
{Aoki}, K., {Kawaguchi}, T., \& {Ohta}, K. 2005, \apj, 618, 601,
  \dodoi{10.1086/426075}

\bibitem[{{Ba{\~n}ados} {et~al.}(2018){Ba{\~n}ados}, {Venemans},
  {Mazzucchelli}, {Farina}, {Walter}, {Wang}, {Decarli}, {Stern}, {Fan},
  {Davies}, {Hennawi}, {Simcoe}, {Turner}, {Rix}, {Yang}, {Kelson}, {Rudie}, \&
  {Winters}}]{Banados18}
{Ba{\~n}ados}, E., {Venemans}, B.~P., {Mazzucchelli}, C., {et~al.} 2018, \nat,
  553, 473, \dodoi{10.1038/nature25180}

\bibitem[{{Bardeen}(1970)}]{Bardeen70}
{Bardeen}, J.~M. 1970, \nat, 226, 64, \dodoi{10.1038/226064a0}

\bibitem[{{Begelman}(1978)}]{Begelman78}
{Begelman}, M.~C. 1978, \mnras, 184, 53, \dodoi{10.1093/mnras/184.1.53}

\bibitem[{{Chon} {et~al.}(2016){Chon}, {Hirano}, {Hosokawa}, \&
  {Yoshida}}]{Chon16}
{Chon}, S., {Hirano}, S., {Hosokawa}, T., \& {Yoshida}, N. 2016, \apj, 832,
  134, \dodoi{10.3847/0004-637X/832/2/134}

\bibitem[{{Collin} {et~al.}(2002){Collin}, {Boisson}, {Mouchet}, {Dumont},
  {Coup{\'e}}, {Porquet}, \& {Rokaki}}]{Collin02}
{Collin}, S., {Boisson}, C., {Mouchet}, M., {et~al.} 2002, \aap, 388, 771,
  \dodoi{10.1051/0004-6361:20020550}

\bibitem[{{Davies} {et~al.}(2019){Davies}, {Hennawi}, \& {Eilers}}]{Davies19}
{Davies}, F.~B., {Hennawi}, J.~F., \& {Eilers}, A.-C. 2019, \apjl, 884, L19,
  \dodoi{10.3847/2041-8213/ab42e3}

\bibitem[{Dong {et~al.}(2012)Dong, Greene, \& Ho}]{Dong12}
Dong, R., Greene, J.~E., \& Ho, L.~C. 2012, The Astrophysical Journal, 761, 73,
  \dodoi{10.1088/0004-637x/761/1/73}

\bibitem[{{Fanidakis} {et~al.}(2012){Fanidakis}, {Baugh}, {Benson}, {Bower},
  {Cole}, {Done}, {Frenk}, {Hickox}, {Lacey}, \& {Del P.~Lagos}}]{Fanidakis12}
{Fanidakis}, N., {Baugh}, C.~M., {Benson}, A.~J., {et~al.} 2012, \mnras, 419,
  2797, \dodoi{10.1111/j.1365-2966.2011.19931.x}

\bibitem[{Hirano {et~al.}(2014)Hirano, Hosokawa, Yoshida, Umeda, Omukai,
  Chiaki, \& Yorke}]{Hirano14}
Hirano, S., Hosokawa, T., Yoshida, N., {et~al.} 2014, The Astrophysical
  Journal, 781, 60, \dodoi{10.1088/0004-637x/781/2/60}

\bibitem[{{Hirschmann} {et~al.}(2012){Hirschmann}, {Somerville}, {Naab}, \&
  {Burkert}}]{Hirschmann12}
{Hirschmann}, M., {Somerville}, R.~S., {Naab}, T., \& {Burkert}, A. 2012,
  \mnras, 426, 237, \dodoi{10.1111/j.1365-2966.2012.21626.x}

\bibitem[{Inayoshi {et~al.}(2016)Inayoshi, Haiman, \& Ostriker}]{Inayoshi16}
Inayoshi, K., Haiman, Z., \& Ostriker, J.~P. 2016, Monthly Notices of the Royal
  Astronomical Society, 459, 3738, \dodoi{10.1093/mnras/stw836}

\bibitem[{{Ishiyama} {et~al.}(2015){Ishiyama}, {Enoki}, {Kobayashi}, {Makiya},
  {Nagashima}, \& {Oogi}}]{Ishiyama15}
{Ishiyama}, T., {Enoki}, M., {Kobayashi}, M.~A.~R., {et~al.} 2015, \pasj, 67,
  61, \dodoi{10.1093/pasj/psv021}

\bibitem[{{Kauffmann} \& {Haehnelt}(2000)}]{KH00}
{Kauffmann}, G., \& {Haehnelt}, M. 2000, \mnras, 311, 576,
  \dodoi{10.1046/j.1365-8711.2000.03077.x}

\bibitem[{{Kawaguchi}(2003)}]{Kawaguchi03}
{Kawaguchi}, T. 2003, \apj, 593, 69, \dodoi{10.1086/376404}

\bibitem[{{Kawaguchi} {et~al.}(2004){Kawaguchi}, {Aoki}, {Ohta}, \&
  {Collin}}]{Kawaguchi04June}
{Kawaguchi}, T., {Aoki}, K., {Ohta}, K., \& {Collin}, S. 2004, \aap, 420, L23,
  \dodoi{10.1051/0004-6361:20040157}

\bibitem[{{Kelly} \& {Shen}(2013)}]{KS13}
{Kelly}, B.~C., \& {Shen}, Y. 2013, \apj, 764, 45,
  \dodoi{10.1088/0004-637X/764/1/45}

\bibitem[{{Komossa} {et~al.}(2008){Komossa}, {Xu}, {Zhou}, {Storchi-Bergmann},
  \& {Binette}}]{Komossa08}
{Komossa}, S., {Xu}, D., {Zhou}, H., {Storchi-Bergmann}, T., \& {Binette}, L.
  2008, \apj, 680, 926, \dodoi{10.1086/587932}

\bibitem[{{Kormendy} \& {Ho}(2013)}]{KH13_review}
{Kormendy}, J., \& {Ho}, L.~C. 2013, \araa, 51, 511,
  \dodoi{10.1146/annurev-astro-082708-101811}

\bibitem[{Liu {et~al.}(2018)Liu, Yuan, Dong, Zhou, \& Liu}]{Liu18}
Liu, H.-Y., Yuan, W., Dong, X.-B., Zhou, H., \& Liu, W.-J. 2018, The
  Astrophysical Journal Supplement Series, 235, 40,
  \dodoi{10.3847/1538-4365/aab88e}

\bibitem[{{Lynden-Bell}(1969)}]{Lynden-Bell69}
{Lynden-Bell}, D. 1969, \nat, 223, 690, \dodoi{10.1038/223690a0}

\bibitem[{{Madau} \& {Rees}(2001)}]{MR01}
{Madau}, P., \& {Rees}, M.~J. 2001, \apjl, 551, L27, \dodoi{10.1086/319848}

\bibitem[{{Magorrian} {et~al.}(1998){Magorrian}, {Tremaine}, {Richstone},
  {Bender}, {Bower}, {Dressler}, {Faber}, {Gebhardt}, {Green}, {Grillmair},
  {Kormendy}, \& {Lauer}}]{Magorrian98}
{Magorrian}, J., {Tremaine}, S., {Richstone}, D., {et~al.} 1998, \aj, 115,
  2285, \dodoi{10.1086/300353}

\bibitem[{{Makiya} {et~al.}(2016){Makiya}, {Enoki}, {Ishiyama}, {Kobayashi},
  {Nagashima}, {Okamoto}, {Okoshi}, {Oogi}, \& {Shirakata}}]{nu2gc}
{Makiya}, R., {Enoki}, M., {Ishiyama}, T., {et~al.} 2016, \pasj, 68, 25,
  \dodoi{10.1093/pasj/psw005}

\bibitem[{{Marconi} {et~al.}(2004){Marconi}, {Risaliti}, {Gilli}, {Hunt},
  {Maiolino}, \& {Salvati}}]{Marconi04}
{Marconi}, A., {Risaliti}, G., {Gilli}, R., {et~al.} 2004, \mnras, 351, 169,
  \dodoi{10.1111/j.1365-2966.2004.07765.x}

\bibitem[{{McLure} \& {Dunlop}(2004)}]{MD04}
{McLure}, R.~J., \& {Dunlop}, J.~S. 2004, \mnras, 352, 1390,
  \dodoi{10.1111/j.1365-2966.2004.08034.x}

\bibitem[{{Menci} {et~al.}(2014){Menci}, {Gatti}, {Fiore}, \&
  {Lamastra}}]{Menci14}
{Menci}, N., {Gatti}, M., {Fiore}, F., \& {Lamastra}, A. 2014, \aap, 569, A37,
  \dodoi{10.1051/0004-6361/201424217}

\bibitem[{{Mineshige} {et~al.}(2000){Mineshige}, {Kawaguchi}, {Takeuchi}, \&
  {Hayashida}}]{MK00}
{Mineshige}, S., {Kawaguchi}, T., {Takeuchi}, M., \& {Hayashida}, K. 2000,
  \pasj, 52, 499, \dodoi{10.1093/pasj/52.3.499}

\bibitem[{{Mortlock} {et~al.}(2011){Mortlock}, {Warren}, {Venemans}, {Patel},
  {Hewett}, {McMahon}, {Simpson}, {Theuns}, {Gonz{\'a}les-Solares}, {Adamson},
  {Dye}, {Hambly}, {Hirst}, {Irwin}, {Kuiper}, {Lawrence}, \&
  {R{\"o}ttgering}}]{Mortlock11}
{Mortlock}, D.~J., {Warren}, S.~J., {Venemans}, B.~P., {et~al.} 2011, \nat,
  474, 616, \dodoi{10.1038/nature10159}

\bibitem[{{Mutlu-Pakdil} {et~al.}(2016){Mutlu-Pakdil}, {Seigar}, \&
  {Davis}}]{MSD16}
{Mutlu-Pakdil}, B., {Seigar}, M.~S., \& {Davis}, B.~L. 2016, \apj, 830, 117,
  \dodoi{10.3847/0004-637X/830/2/117}

\bibitem[{{Nobuta} {et~al.}(2012){Nobuta}, {Akiyama}, {Ueda}, {Watson},
  {Silverman}, {Hiroi}, {Ohta}, {Iwamuro}, {Yabe}, {Tamura}, {Moritani},
  {Sumiyoshi}, {Takato}, {Kimura}, {Maihara}, {Dalton}, {Lewis}, {Bonfield},
  {Lee}, {Curtis-Lake}, {Macaulay}, {Clarke}, {Sekiguchi}, {Simpson}, {Croom},
  {Ouchi}, {Hanami}, \& {Yamada}}]{Nobuta12}
{Nobuta}, K., {Akiyama}, M., {Ueda}, Y., {et~al.} 2012, \apj, 761, 143,
  \dodoi{10.1088/0004-637X/761/2/143}

\bibitem[{{Novak}(2013)}]{Novak13}
{Novak}, G.~S. 2013, arXiv e-prints.
\newblock \doarXiv{1310.3833}

\bibitem[{{Omukai} {et~al.}(2008){Omukai}, {Schneider}, \& {Haiman}}]{OSH08}
{Omukai}, K., {Schneider}, R., \& {Haiman}, Z. 2008, \apj, 686, 801,
  \dodoi{10.1086/591636}

\bibitem[{{Pezzulli} {et~al.}(2016){Pezzulli}, {Valiante}, \&
  {Schneider}}]{PVS16}
{Pezzulli}, E., {Valiante}, R., \& {Schneider}, R. 2016, \mnras, 458, 3047,
  \dodoi{10.1093/mnras/stw505}

\bibitem[{{Regan} {et~al.}(2019){Regan}, {Downes}, {Volonteri}, {Beckmann},
  {Lupi}, {Trebitsch}, \& {Dubois}}]{Regan19}
{Regan}, J.~A., {Downes}, T.~P., {Volonteri}, M., {et~al.} 2019, \mnras, 486,
  3892, \dodoi{10.1093/mnras/stz1045}

\bibitem[{{Salpeter}(1964)}]{Salpeter64}
{Salpeter}, E.~E. 1964, \apj, 140, 796, \dodoi{10.1086/147973}

\bibitem[{{Schulze} \& {Wisotzki}(2010)}]{SW10}
{Schulze}, A., \& {Wisotzki}, L. 2010, \aap, 516, A87,
  \dodoi{10.1051/0004-6361/201014193}

\bibitem[{{Shakura} \& {Sunyaev}(1973)}]{SS73}
{Shakura}, N.~I., \& {Sunyaev}, R.~A. 1973, \aap, 24, 337

\bibitem[{{Shankar} {et~al.}(2004){Shankar}, {Salucci}, {Granato}, {De Zotti},
  \& {Danese}}]{Shankar04}
{Shankar}, F., {Salucci}, P., {Granato}, G.~L., {De Zotti}, G., \& {Danese}, L.
  2004, \mnras, 354, 1020, \dodoi{10.1111/j.1365-2966.2004.08261.x}

\bibitem[{{Shankar} {et~al.}(2009){Shankar}, {Weinberg}, \&
  {Miralda-Escud{\'e}}}]{SWM09}
{Shankar}, F., {Weinberg}, D.~H., \& {Miralda-Escud{\'e}}, J. 2009, \apj, 690,
  20, \dodoi{10.1088/0004-637X/690/1/20}

\bibitem[{Shirakata {et~al.}(2019b)Shirakata, Kawaguchi, Oogi, Okamoto, \&
  Nagashima}]{Shirakata19b}
Shirakata, H., Kawaguchi, T., Oogi, T., Okamoto, T., \& Nagashima, M. 2019b,
  Monthly Notices of the Royal Astronomical Society, 487, 409,
  \dodoi{10.1093/mnras/stz1282}

\bibitem[{{Shirakata} {et~al.}(2016){Shirakata}, {Kawaguchi}, {Okamoto},
  {Makiya}, {Ishiyama}, {Matsuoka}, {Nagashima}, {Enoki}, {Oogi}, \&
  {Kobayashi}}]{Shirakata16}
{Shirakata}, H., {Kawaguchi}, T., {Okamoto}, T., {et~al.} 2016, \mnras, 461,
  4389, \dodoi{10.1093/mnras/stw1798}

\bibitem[{{Shirakata} {et~al.}(2019a){Shirakata}, {Okamoto}, {Kawaguchi},
  {Nagashima}, {Ishiyama}, {Makiya}, {Kobayashi}, {Enoki}, {Oogi}, \&
  {Okoshi}}]{Shirakata19}
{Shirakata}, H., {Okamoto}, T., {Kawaguchi}, T., {et~al.} 2019a, \mnras, 482,
  4846, \dodoi{10.1093/mnras/sty2958}

\bibitem[{{So\l tan}(1982)}]{Soltan82}
{So\l tan}, A. 1982, \mnras, 200, 115

\bibitem[{Tanaka \& Haiman(2009)}]{TH09}
Tanaka, T., \& Haiman, Z. 2009, The Astrophysical Journal, 696, 1798,
  \dodoi{10.1088/0004-637x/696/2/1798}

\bibitem[{{Tucci} \& {Volonteri}(2017)}]{TV17}
{Tucci}, M., \& {Volonteri}, M. 2017, \aap, 600, A64,
  \dodoi{10.1051/0004-6361/201628419}

\bibitem[{{Valiante} {et~al.}(2016){Valiante}, {Schneider}, {Volonteri}, \&
  {Omukai}}]{Valiante16J}
{Valiante}, R., {Schneider}, R., {Volonteri}, M., \& {Omukai}, K. 2016, \mnras,
  457, 3356, \dodoi{10.1093/mnras/stw225}

\bibitem[{{Vika} {et~al.}(2009){Vika}, {Driver}, {Graham}, \& {Liske}}]{Vika09}
{Vika}, M., {Driver}, S.~P., {Graham}, A.~W., \& {Liske}, J. 2009, \mnras, 400,
  1451, \dodoi{10.1111/j.1365-2966.2009.15544.x}

\bibitem[{{Watarai} {et~al.}(2000){Watarai}, {Fukue}, {Takeuchi}, \&
  {Mineshige}}]{Watarai00}
{Watarai}, K.-y., {Fukue}, J., {Takeuchi}, M., \& {Mineshige}, S. 2000, \pasj,
  52, 133, \dodoi{10.1093/pasj/52.1.133}

\bibitem[{{Wu} {et~al.}(2015){Wu}, {Wang}, {Fan}, {Yi}, {Zuo}, {Bian}, {Jiang},
  {McGreer}, {Wang}, {Yang}, {Yang}, {Thompson}, \& {Beletsky}}]{Wu15}
{Wu}, X.-B., {Wang}, F., {Fan}, X., {et~al.} 2015, \nat, 518, 512,
  \dodoi{10.1038/nature14241}

\bibitem[{{Yu} \& {Tremaine}(2002)}]{YT02}
{Yu}, Q., \& {Tremaine}, S. 2002, \mnras, 335, 965,
  \dodoi{10.1046/j.1365-8711.2002.05532.x}

\bibitem[{{Zamanov} {et~al.}(2002){Zamanov}, {Marziani}, {Sulentic}, {Calvani},
  {Dultzin-Hacyan}, \& {Bachev}}]{Zamanov02}
{Zamanov}, R., {Marziani}, P., {Sulentic}, J.~W., {et~al.} 2002, \apjl, 576,
  L9, \dodoi{10.1086/342783}

\end{thebibliography}

\end{document}